\documentclass[apjl]{emulateapj}
\input{psfig.sty}
\bibpunct{(}{)}{;}{a}{}{,}

\newcommand{\n}{\newcommand}
\n{\bmain}{\begin{list}{${\bullet}$}{}}
\n{\emain}{\end{list}}
\n{\bminor}{\begin{list}{$\bf\triangleright$}{}}
\n{\eminor}{\end{list}}
\n{\Ms}{\mbox{$\,M_\odot$}}

\shorttitle{Dichotomous Neutron-Star Kicks}
\shortauthors{Podsiadlowski et al.}

\begin{document}

\title{The Effects of Binary Evolution on the Dynamics of Core Collapse and 
       Neutron-Star Kicks}


\author{Ph.\ Podsiadlowski\altaffilmark{1}, N.\ Langer\altaffilmark{2}, 
A. J. T. Poelarends\altaffilmark{2}, S. Rappaport\altaffilmark{3}, A. 
Heger\altaffilmark{4}, and E. Pfahl\altaffilmark{5}}

\altaffiltext{1}{Department of Astrophysics, University of Oxford,
Oxford, OX1 3RH, UK, podsi@astro.ox.ac.uk}

\altaffiltext{2}{Astronomical Institute, P.O. Box 80000, NL-3508 TA
Utrecht, The Netherlands, n.langer@astro.uu.nl}

\altaffiltext{3}{Department of Physics and Center for Space Research,
Massachusetts Institute of Technology, sar@mit.edu}

\altaffiltext{4}{Department of Astronomy and Astrophysics, University
of Chicago, 5640 S. Ellis Avenue, Chicago, IL 60637, USA, and
Theoretical Astrophysics Group, T-6, MS B227, Los Alamos National
Laboratory, Los Alamos, NM 87545, USA, alex@6.lanl.gov}

\altaffiltext{5}{Chandra Fellow, Harvard-Smithonian Center for
Astrophysics, 60 Garden Street, Cambridge, MA 02138, USA,
epfahl@cfa.harvard.edu}




\begin{abstract}
We systematically examine how the presence in a binary affects the
final core structure of a massive star and its consequences for the
subsequent supernova explosion. Interactions with a companion star may
change the final rate of rotation, the size of the helium core, the
strength of carbon burning and the final iron core mass. Stars with
initial masses larger than $\sim 11\Ms$ that experience core collapse
will generally have smaller iron cores at the point of explosion if
they lost their envelopes due to a binary interaction during or soon
after core hydrogen burning. Stars below $\sim 11 M_{\odot}$, on the
other hand, can end up with larger helium and metal cores if they have
a close companion, since the second dredge-up phase which reduces the
helium core mass dramatically in single stars does not occur once the
hydrogen envelope is lost. We find that the initially more massive
stars in binary systems with masses in the range $8 - 11 M_{\odot}$
are likely to undergo an electron-capture supernova, while single
stars in the same mass range would end as ONeMg white dwarfs.  We
suggest that the core collapse in an electron-capture supernova (and
possibly in the case of relatively small iron cores) leads to a prompt
or fast explosion rather than a very slow, delayed neutrino-driven
explosion and that this naturally produces neutron stars with
low-velocity kicks. This leads to a dichotomous distribution of
neutron star kicks, as inferred previously, where neutron stars in
relatively close binaries attain low kick velocities. We illustrate
the consequences of such a dichotomous kick scenario using binary
population synthesis simulations and discuss its implications. This
scenario has also important consequences for the minimum initial mass
of a massive star that becomes a neutron star.  For single stars the
critical mass may be as high as 10\,--\,12\Ms, while for close
binaries, it may be as low as 6\,--\,8\Ms. These critical masses
depend on the treatment of convection, the amount of convective
overshooting and the metallicity of the star and will generally be
lower for larger amounts of convective overshooting and lower
metallicity.
\end{abstract}



\section{Introduction}
For the last decade, it has generally been accepted that the high
speeds inferred for isolated, normal radio pulsars imply that neutron
stars (NSs) receive a large impulse, or ``kick,'' at birth.  Measured
proper motions for $\simeq$100 radio pulsars indicate typical kick
speeds in excess of 100\,--\,200\,km\,s$^{-1}$
\citep{Lyne1994,Hansen1997,Cordes1998,Arzoumanian2002}, though the
functional form of the underlying speed distribution is poorly
constrained.  The physical mechanism that causes this kick is
presently unknown, but is presumably the result of some asymmetry in
the core collapse or subsequent supernova (SN) explosion (see, e.g.,
\citet{Pfahl2002c} for discussion and references).

In apparent conflict with the high speeds of isolated radio pulsars,
\citet{Pfahl2002c} identified a new class of high-mass X-ray binaries
(HMXBs) wherein the neutron stars must have been born with relatively
low kick speeds.  These HMXBs are distinguished by their long orbital
periods ($P_{\rm orb}> 30\,$d) and low eccentricities ($e\la 0.2$),
with the prime example being X Per/4U~0352+309 ($P_{\rm orb}\simeq
250\,$d; $e \simeq 0.1$).  The orbits of these systems are
sufficiently wide that tidal circularization is negligible, so that
the observed eccentricities should reflect the conditions immediately
after the SN explosion.  Such low eccentricities essentially require
that the neutron stars in these systems received low kick speeds of
$\la$$50\,$km\,s$^{-1}$.

Further observational evidence that at least some neutron stars
receive low kicks at birth is provided by the fact that a large number
of neutron stars are found in globular clusters, where some massive
globular clusters may contain more than $\sim 1000$ neutron stars
\citep{Pfahl2002b}.  Since the central escape velocity is generally
$\la 50\,$km\,s$^{-1}$, essentially all of the neutron stars born in a
globular cluster should escape from the cluster if they received a
kick consistent with the kick distribution for single radio pulsars
(for a detailed discussion of this so-called `neutron-star retention
problem' see Pfahl et al.\ 2002b).  If all of these neutron stars
were originally born in massive binaries, this would alleviate the
problem somewhat, since in this case the momentum imparted to the
neutron star would be shared by the total mass of the system, leading
to a correspondingly smaller space velocity of the post-supernova
binary \citep{Brandt1995}. However, while the membership in a binary
increases the number of neutron stars that can remain bound in the
cluster significantly, the effect may not be large enough to explain the
observed numbers unless clusters were much more massive at an earlier
epoch than they are today (Drukier 1996; Pfahl et al.\ 2002b).

To simultaneously account for the new class of HMXBs and the high
speeds of radio pulsars, \citet{Pfahl2002c} suggested that neutron
stars originating from progenitors that are single or members of wide
binaries receive the conventional large kicks, while neutron stars
born in close binaries receive small kicks (also see Katz 1975, 1983
and Hartman 1997 for earlier more {\it ad hoc} suggestions of a
significant population of neutron stars with small natal kicks).
\citet{Pfahl2002c} further argued that the proposed dichotomy between
high and low kick speeds, and its relation to the evolutionary history
of the NS progenitor, might be associated with the rotation rate of
the collapsing core.  However, as we will show in this paper, the core
structure itself (e.g., its mass and composition) depends strongly on
whether a star evolves in a close binary or in isolation. This can
produce differences in the actual supernova and may allow a prompt (or
at least a fast) supernova explosion mechanism to be successful in
some cases; this in turn may lead to relatively low supernova kicks.

%

The outline of this paper is the following: In \S~2 we systematically
discuss the differences in the core evolution for stars in close
binaries and in single systems/wide binaries and discuss the
implications for the core-collapse phase. In \S~3 we develop a general
scheme for forming neutron stars in different types of systems and the
expected differences in kick velocity, associating them with
individual known systems or classes of systems. In \S~4 we discuss how
this scheme can be tested both observationally and suggest how further
progress can be made on the theoretical side.

%
\section{Binary evolution and the pre-core-collapse core structure
of massive stars}

It is often naively assumed that the evolution of helium cores is the
same irrespective of whether the core is surrounded by a hydrogen
envelope or not, and that the final core structure will be similar in
the two cases. However, binary evolution may affect the final
pre-supernova structure of a massive star in several fundamentally
different ways: (1) the rate of rotation of the immediate
pre-supernova core, (2) the size of the helium core, (3) the
occurrence of a second dredge-up phase at the beginning of the
asymptotic-giant branch (AGB), and (4) the C/O ratio at the end of
helium burning which affects the strength of carbon burning and the
final size of the iron core. These effects can dramatically change the
condition of core collapse as first pointed out by \citet{Brown1999}
and in particular \citet{Brown2001}, who showed that the final iron
cores of massive stars will be significantly smaller for stars that
have lost their hydrogen-rich envelopes soon after the end of core
hydrogen burning.

\bigskip
\noindent{\em The role of rotation}
\medskip

While massive stars are generally rapid rotators on the main sequence,
there are several efficient mechanisms by which they lose their
angular momentum during their evolution.  The final rotation rate of
the core of a massive star depends on whether the star passed through
a red-supergiant phase during which the core will be effectively
braked by the hydrodynamic and magnetic coupling to the slowly
rotating envelope \citep{Spruit1998,Heger2003} and the wind mass loss
during a subsequent Wolf-Rayet/helium-star phase, which can be very
efficient in extracting angular momentum from a helium star
\citep{Heger2003b} and slowing it down in the process.
\citet{Pfahl2002c} argued that stars that lost their H-rich envelopes
soon after the main-sequence phase (so-called early Case B mass
transfer\footnote{It is common practice to distinguish among three
evolutionary phases of the primary at the onset of mass transfer,
following \citet{Kippenhahn1967} (see also
\citealt{Lauterborn1970,Podsiadlowski1992}). Case A evolution
corresponds to core hydrogen burning, Case B refers to the shell
hydrogen-burning phase, but prior to central helium ignition, and Case
C evolution begins after helium has been depleted in the core.}) might
avoid this phase where the core is effectively braked and may still be
rapidly rotating at the time of core collapse.

However, whether the star can avoid spin-down also depends on whether
mass transfer is dynamically stable or unstable. In the case of stable
Case B Roche-lobe overflow, \citet{Langer2003a} showed that, if the
mass-losing star remains tidally locked to the orbit, this provides a
very efficient method of slowing down the rotation rate of the
mass-losing star. On the other hand, in the case of unstable mass
transfer, leading to a common-envelope and spiral-in phase
\citep{Paczynski1976}, the spiral-in timescale is much shorter than
any realistic synchronization timescale, and tidal locking would not
be expected. In this case, the core of the mass-losing star could
still be rapidly rotating after the ejection of the common
envelope. This implies that the scenario suggested by
\citet{Pfahl2002c} probably requires late case B mass transfer (i.e.,
dynamically unstable mass transfer that leads to a CE phase before the
core has been spun down significantly).

Another situation which may lead to a rapidly rotating core is the
complete slow merger of two massive stars, in particular if it occurs
after helium core burning, as in models for the progenitor of SN 1987A
\citep{Ivanova2003} which predict a very rapidly rotating core for the
immediate pre-supernova star (also see Joss \& Becker 2003).

Finally we note that, if the exploding star is still accreting from a
companion at the time of the supernova, one would also expected a
rapidly rotating core \citep{Langer2003b}.

As this discussion shows, binary interactions can significantly affect
the final pre-supernova rotation rate of the core of a massive star,
although the details can be rather involved and are not completely
understood. In addition, what is even less clear at the present time
is how this affects the physics of the core collapse itself and, in
particular, the magnitude of the kick imparted to the newborn neutron
star.

\bigskip
\noindent{\em The size of the helium core and the second dredge-up}
\medskip

The final size of the helium core depends strongly on the evolution of
the H-rich envelope. During core helium burning, the helium core
ordinarily grows substantially because of hydrogen burning in a shell
-- often the dominant nuclear burning source -- which adds fresh
helium to the core. On the other hand, in a binary a star may lose its
H-rich envelope before He burning (or early during the He-burning
phase). In this case, the helium core can no longer grow and may in
fact shrink because of the strong stellar wind expected in the
subsequent Wolf-Rayet phase (e.g., 
\citealt{Woosley1995a,Wellstein1999,Wellstein2001,Pols2002}).
Therefore, the final mass of the helium core will often be lower for
stars in close binaries than in single systems/wide binaries (see
Fig.~1).

Another factor that strongly affects the mass of the final helium core
is the occurrence of a second dredge-up phase. When stars up to $\sim
11\Ms$ ascend the asymptotic giant branch, they generally experience a
second dredge-up phase where the convective envelope penetrates deep
into the H-exhausted core and dredges up a significant fraction of the
helium core \citep{Iben1974}. As a consequence, the size of the helium
core can be dramatically reduced (by up to $\sim 1.6\Ms$; see Fig.~1).
However, the occurrence of the second dredge-up depends on the
presence of a convective H-rich envelope. If the star loses its H-rich
envelope before this phase, dredge-up does not occur. In this case,
the final size of the helium core is larger for a star in a binary
than its single counterpart (note that this is the opposite of what
happens to their more massive counterparts).  This is illustrated in
Figure~1 which shows the final helium core mass as a function of
initial main-sequence mass for single stars (thick solid curve;
\citealt{Poelarends2004}) and for stars in binaries that lose their
envelopes either in Case A or Case B mass transfer (based on the
results of \citealt{Wellstein2001}). The almost discontinuous change
of the final helium mass around 12\Ms\ is a direct consequence of the
fact that stars below this mass have experienced a second dredge-up
phase, while stars above this mass do not or only dredge up a moderate
amount of the helium core.  Note that the final mass for these stars
is less than the minimum mass for core collapse ($\sim 1.4\Ms$). After
the dredge-up phase, the helium core may grow again because of
hydrogen shell burning, just as in a normal AGB star. Whether the core
can reach the critical mass for core collapse depends on the timescale
on which the star loses its envelope in a stellar wind or
superwind. While this is somewhat uncertain, we estimate that for the
models in Figure~1, single stars below $\sim 12\Ms$ will not reach the
critical mass for core collapse and will end their evolution as ONeMg
white dwarfs. There may be a small mass range around this boundary
where single stars reach the condition for core collapse.

In contrast, the shaded region between the dot-dashed thick lines
indicates the expected range of the final helium core mass for stars
that lose their envelopes by binary interactions (in this case, the
final core mass depends on the evolutionary phase and the core mass at
the time of mass transfer).

In the past, \citet{Nomoto1984,Nomoto1987a} has argued that an
electron-capture supernova is the expected fate for stars with
main-sequence masses in the range of 8\,--\,10\Ms. In his
calculations, these stars developed helium cores in the range of
$M=2.0$\,--\,2.5\Ms\footnote{We note that in Nomoto's calculations,
stars in the range of 8\,--\,10\Ms\ either did not experience a second
dredge-up phase or only during carbon shell burning. As a consequence, most
of his models in this mass range, unlike ours, experienced a
core-collapse supernova even in the case of single stars. We suspect
that this difference can be attributed to the different opacities
employed. The new OPAL opacities are significantly larger in the
critical temperature range, which makes dredge-up more likely or occur
earlier. We emphasize that this dredge-up behaviour is qualitatively
found in all other recent, detailed studies of stars in this mass
range \citep{Ritossa1996,Berro1997,Iben1997,Eldridge2004}. We note,
however, that the range of initial masses which leads to helium cores
in the range of 2.0\,--\,2.5\Ms\ depends on some of the uncertainties
in the stellar modelling, in particular the treatment of convection
and convective overshooting (for further discussion of the
uncertainties see \S~2.2).} and never developed an iron core; in this
case, the collapse is triggered by the electron capture on $^{24}$Mg
and $^{20}$Ne \citep{Nomoto1984,Nomoto1987a} (for a recent discussion
see \citealt{Wanajo2003}). In Figure~1, the light dashed horizontal
lines indicate approximately the range of helium-core masses which
can be expected to lead to an electron-capture supernova (based on the
results of \citet{Nomoto1984,Nomoto1987a}; also see
\citealt{Habets1986}).  As Figure~1 shows, because of the second
dredge-up, the mass range for which single stars experience an
electron-capture supernova may be very narrow if non-existent, while
there is a large mass range for which it may occur for a star in a
close binary.  Indeed a binary channel may be the only place where it
can occur.

\bigskip
\noindent{\em The C/O ratio and the strength of carbon burning}
\medskip

Another more subtle effect is that the lack of a H-burning shell
leads to a lower C/O ratio at the end of helium core burning
which affects the strength of subsequent carbon burning and the final
size of the iron core. This effect was first pointed out and
explained by \citet{Brown2001} and can be understood as follows.

\begin{figure}
\centerline{\psfig{file=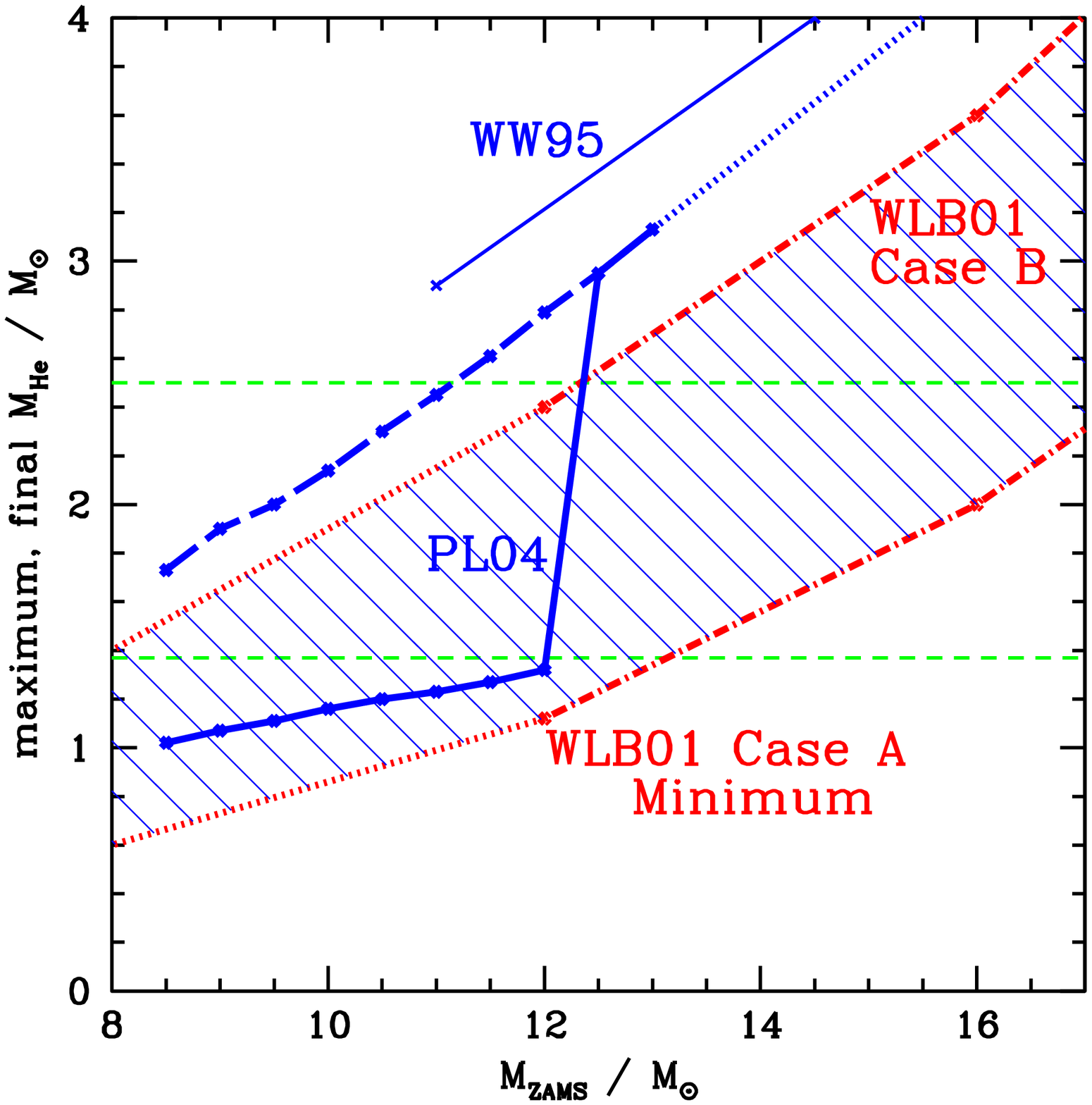,width=8cm}}
\caption{Final mass (thick solid line) and maximum mass (thick dashed
line) of the helium core in single stars as a function of initial mass
according to \citet{Poelarends2004} (PL04), extrapolated for initial
masses above 12.5\Ms. The final helium core masses from the
calculations of \citet{Woosley1995b} are indicated by a thin solid
line. The final helium core masses of close binary models undergoing
Case B mass transfer from \citet{Wellstein2001} (WLB01) are shown as
the upper dot-dashed line, while those experiencing Case A mass
transfer (WLB01) may lie anywhere between the lower dot-dashed line and
the Case B line. The results from the binary calculations have been
extrapolated for initial masses below 12\Ms\ (dotted part). Note
that the PL04 and WLB01 models have been computed with the same
assumptions for convective mixing, while the WW95 models assumed a
higher semiconvective mixing efficiency. The light
dashed horizontal lines give the range for the final helium core mass for
which the star may experience an electron-capture supernova.
Note that the parameter range for which this may occur
for a single star is very small.}
\label{fig1}
\end{figure}
In the late phase of helium core burning, the alpha-capture reaction
C$^{12}+\alpha\rightarrow$ O$^{16}$, which destroys carbon, tends to
dominate over the carbon-producing triple-$\alpha$ reaction (because
of the different functional dependence on the number density of
$\alpha$-particles). As a consequence, carbon is systematically being
destroyed in this late core-helium-burning phase (this switch occurs
when the helium mass fraction falls below $Y=0.1\,$--\,0.2). In helium
cores surrounded by a hydrogen-burning shell, both the
hydrogen-exhausted core and the convective helium-burning core
continue to increase during helium core burning. This leads to the
continued injection of fresh helium into the helium-burning core and
prolongs the phase in which carbon is preferentially destroyed. This
eventually produces a smaller carbon abundance at the end of core
helium burning than would be the case for a naked helium core, where
the convective core does not grow. This has drastic consequences for
the subsequent carbon-burning phase if helium or carbon-oxygen cores
of similar sizes are compared.  For high carbon abundances (as
expected for naked helium cores), the phase of convective carbon shell
burning lasts longer, typically leading to smaller carbon-exhausted
cores). This in turn produces smaller iron cores with steeper density
gradients outside the iron core \citep{Fryer2002,Heger2002}.  Thus,
for the same initial size of the helium or carbon core, the higher
carbon abundance in a star that lost its hydrogen-rich envelope before
the central helium abundance dropped below $\sim10\,\%$ will result in
a pre-supernova structure that more easily produces a successful
supernova.

\citet{Brown2001} demonstrated that this dichotomy leads to much
smaller iron cores for massive stars and that, as a consequence, even a
60\,$M_{\odot}$ main-sequence star may produce a neutron star rather
than a black hole if it has lost its envelope before its helium 
core-burning phase. In contrast, the minimum mass for black-hole formation
for single stars may be as low as 20\,$M_{\odot}$ \citep{Fryer1999,Fryer2001}.

In analogy to these results, it is reasonable to expect that there
will be significant differences in the core properties even for massive
stars with $M\la 20\,M_{\odot}$, possibly allowing for successful
prompt (fast) supernova explosions (see e.g., \citealt{Sumiyoshi2001} and
\S~2.1).

\subsection{Prompt (fast) explosions and supernova kicks}

At present, neither the mechanism that produces a successful
core-collapse supernova nor the origin of supernova kicks is properly
understood (see \citealt{Janka2002,Fryer2003} for detailed recent reviews, 
and also Fryer \& Warren (2002, 2003)). In one of the most popular explosion
scenarios, it is the delayed heating by neutrinos that revives the
outgoing supernova shock several 100\,ms after it stalled in the initial
core bounce (i.e., after many dynamical timescales).  In this
scenario, the origin of the supernova kick may be connected with
asymmetries in this long phase where the explosion develops (e.g., due
to the continued accretion onto the proto-neutron star or caused by
the strong convection in the neutrino-heated region;
e.g., \citealt{Janka2002}). In contrast, in a prompt supernova
explosion, the initial bounce drives a successful supernova shock on
the dynamical timescale of the proto-neutron star.  The absence of a
long phase where the explosion teeters at the brink of success could
then be the cause for the absence of a large supernova kick.

Such a scenario is supported by some of the most recent
core-collapse simulations by Scheck et al.\ (2004) (see, however, also
Fryer \& Warren 2003 for a very different view). In the simulations by
Scheck et al.\ (2004), a large supernova kick (up to and above
$500\,{\rm km\,s}^{-1}$) is caused by asymmetries in the neutrino
flux, which have their origin in low-order instabilities driven by the
convective motion behind the stalled shock. An essential requirement
in these simulations, which allows these convective instabilities to
grow, is that the duration of the convective, stalled phase is longer
than $\sim 500\,$ms (i.e., many convective turnover timescales). Such
{\em slow} explosions are expected for fairly large iron cores. In
contrast, for a small iron cores or in the case of electron-capture
supernovae, the simulations by Scheck et al.\ (2004) suggest {\em
fast} explosions where these convectively driven instabilities are
unable to grow, leading to rather small kick velocities\footnote{Note
that, unlike the case of a {\em prompt} explosion, the fast explosions
in the simulations by Scheck et al.\ (2004) occur on a timescale long
compared to the dynamical timescale of the proto-neutron star.}.
 
It seems quite attractive to relate the dichotomy in the supernova
kicks to the differences between a slow and a prompt (fast) supernova
explosion. Whether a supernova explosion develops promptly or in a
delayed manner depends mainly on the difference between the mass of the
Fe-Ni core and the mass of the collapsing core (which in turn also
depends on the initial entropy in the core), since this determines the
amount of shock energy that is consumed in the nuclear dissociation of
heavy elements (see \citet{Hillebrandt1984} for a detailed discussion
and references; \citealt{Sumiyoshi2001}). 

Electron-capture supernovae provide a particularly promising scenario
for a prompt (fast) explosion, since the whole core collapses to
nuclear densities; this makes it much easier for the shock to eject
the envelope, preventing the growth of the instabilities that lead to
large kicks. As our previous discussion shows, electron-capture
supernovae may only (or mainly) occur in close binaries; in this case,
neutron stars with low kicks may be (almost) exclusively produced in
close binary systems (with orbital periods $\la$ a few 100\,d).

\subsection{The minimum mass for core collapse}

An important related issue is the question of the minimum initial
mass, $M_{\rm min}$, of a star that leads to a core-collapse supernova
in a single or binary system.  It is commonly assumed that this
minimum mass is around 8\Ms, the minimum mass above which ignite
carbon off-center (rather than explosively in the center) and form an
ONeMg core \citep{Iben1974}. If the ONeMg core is able to grow to
reach the Chandrasekhar mass, it will collapse in an electron-capture
supernova.  However, if a star experiences mass-transfer already on
the main sequence, an initial star as massive as $\sim 16\Ms$ may end
its evolution as a white dwarf rather than experience core
collapse\footnote{This occurs when the system experiences an
additional RLOF phase after core He exhaustion (Case ABB mass
transfer; \citealt{Wellstein2001,Podsiadlowski2003}) during a He
red-supergiant (RSG) phase; strong mass loss during the He-RSG stage
may lead to a similar outcome.}. On the other hand, as discussed
above, stars below 12\Ms\ may only experience core collapse if they
have lost their envelopes by binary interactions after their
main-sequence phase but before experiencing dredge-up on the AGB.
This implies that the value of $M_{\rm min}$ may be as high as 12\Ms\
for single stars and as low as 8\Ms\ for relatively close binaries.

The exact value for $M_{\rm min}$ is quite sensitive to the treatment
of convection and, in particular, the amount of convective
overshooting, and the metallicity of the star. The value of $M_{\rm
min}= 12\Ms$ was obtained for models that used the Ledoux criterion
for convection without convective overshooting. We estimate that using
the Schwarzschild criterion would reduce $M_{\rm min}$ to $\sim
11\Ms$ if no convective overshooting is included, and to $\sim 10\Ms$
if the recent empirical calibration of convective overshooting
\citep{Pols1997,Schroeder1997} is adopted (also see
\citealt{Ritossa1999,Eldridge2004}). Similarly, the minimum mass for
off-center carbon ignition may be as low as $\sim 6\Ms$ for models including
convective overshooting (Han 2002, unpublished).

\citet{Han1994} also found that, for low-metallicity ($Z=0.001$)
models without convective overshooting, the minimum mass for
off-center carbon ignition was systematically lower ($\sim 6\Ms$) than
for solar metallicity ($\sim 8\Ms$).

As these discussions illustrate, the initial mass range that leads to
the formation of ONeMg white dwarfs, which also determines the minimum
mass for stars that will experience core collapse, depends both on the
details of binary interactions and on the stellar properties.
Different treatments of convection are expected to lead to initial
mass ranges ranging from [6,9]\Ms\ to [9,12]\Ms. In addition at low
metallicity, these ranges should be systematically shifted towards
lower masses (by perhaps $2\Ms$ for $Z=0.001$;
\citealp{Han1994}). Considering that this is an important parameter in
galactic modelling, a thorough re-examination of this issue is
urgently needed \citep{Poelarends2004,Eldridge2004}.

We emphasize that the exact values of this mass range do
not affect the main arguments in this paper, since this only shifts
the mass range in which an electron-capture supernova can be expected,
but does not change the expected dichotomous behaviour.

\section{A dichotomous kick scheme}

The scenario for NS kicks proposed herein has a significant impact on
the theoretical production probabilities and distributions of orbital
parameters of binary systems containing neutron stars.  We illustrate
this by means of a Monte Carlo binary population synthesis (BPS)
calculation. Below we provide a brief description of the important
elements of the code; further details may be found in
\citet{Pfahl2002a,Pfahl2002b,Pfahl2002c,Pfahl2003}. The BPS code
follows a randomly generated sample of massive primordial binaries
through the phase of mass transfer from the primary to the
secondary.\footnote{Here ``primary'' and ``secondary'' refer to the
initially more and less massive star, respectively.}  The initial
primary and secondary masses, $M_{1i}$ and $M_{2i}$, are drawn from
the respective distributions $p(M_{1i}) \propto M_{1i}^{-2.5}$ and
$p(q_i) = 1$, where $q_i\equiv M_{2i}/M_{1i} < 1$ is the initial mass
ratio.  For simplicity, we assume circular orbits, and the initial
orbital separation, $a_i$, is chosen from $p(a_i) \propto a_i^{-1}$.

\begin{figure} 
\begin{center} 
\psfig{file=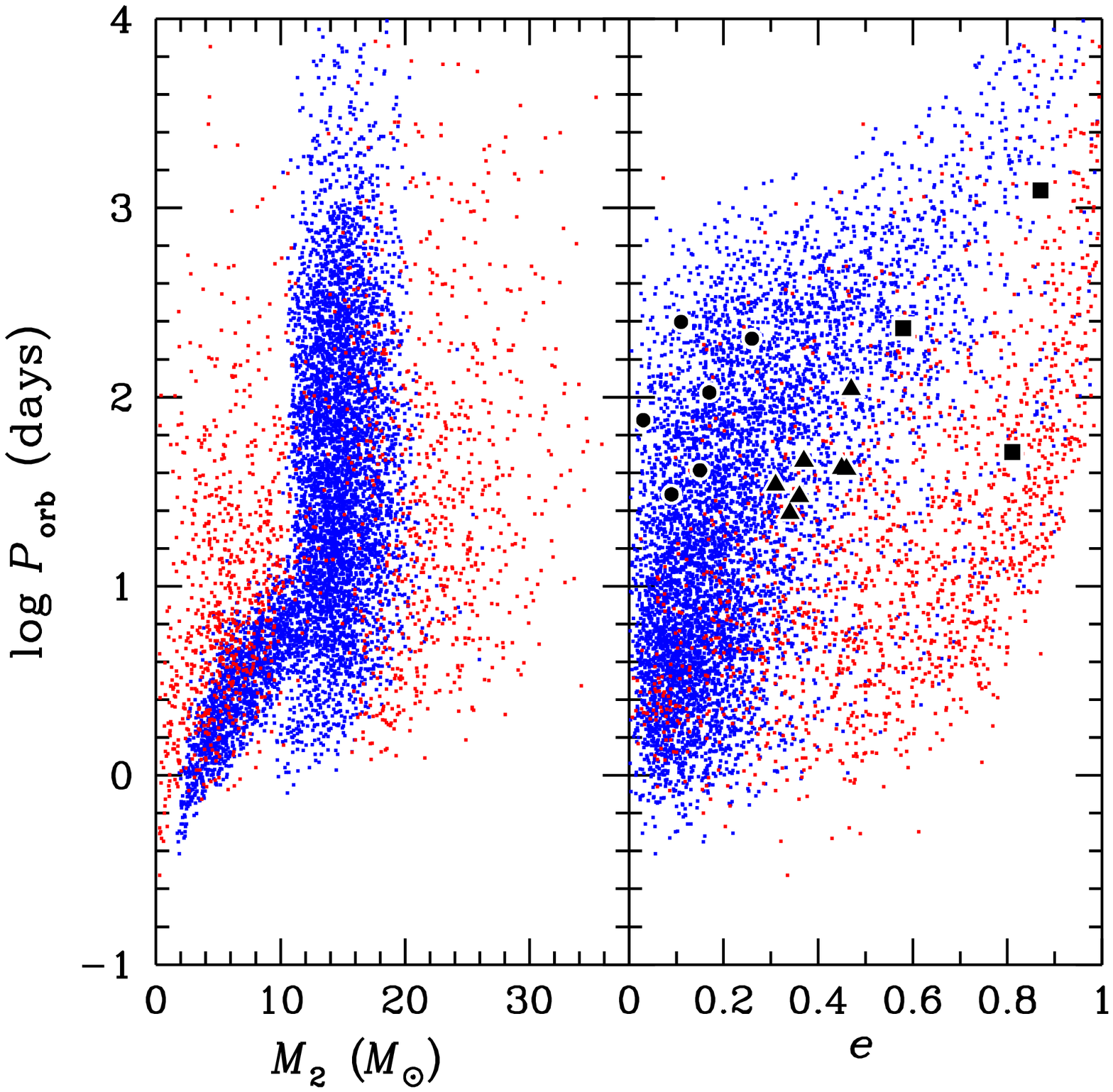,width=0.45\textwidth}
\caption{Results of a single HMXB binary population synthesis
calculation that utilizes our proposed dichotomous kick scenario.
Each red (blue) dot represents a system where the primary was (was
not) highly evolved when it lost its hydrogen-rich envelope, and the
exposed core evolved to collapse to form a neutron star with a
subsequent large, conventional (small, unconventional) natal kick; see
the text for details.  Markers in the right panel indicate the
observed wide, low-eccentricity HMXBs ({\em filled circles}), and the
well-known eccentric HMXBs ({\em filled triangles}).  The {\em filled
squares} show the three radio pulsars with massive binary companions
in eccentric orbits. We did not include observed systems with $P_{\rm
orb} \la 10$ days, since their orbital parameters are likely to have
been altered by tidal circularization effects.\label{fig:BPS}}
\end{center}
\end{figure}

\begin{figure}  
\begin{center}
\psfig{file=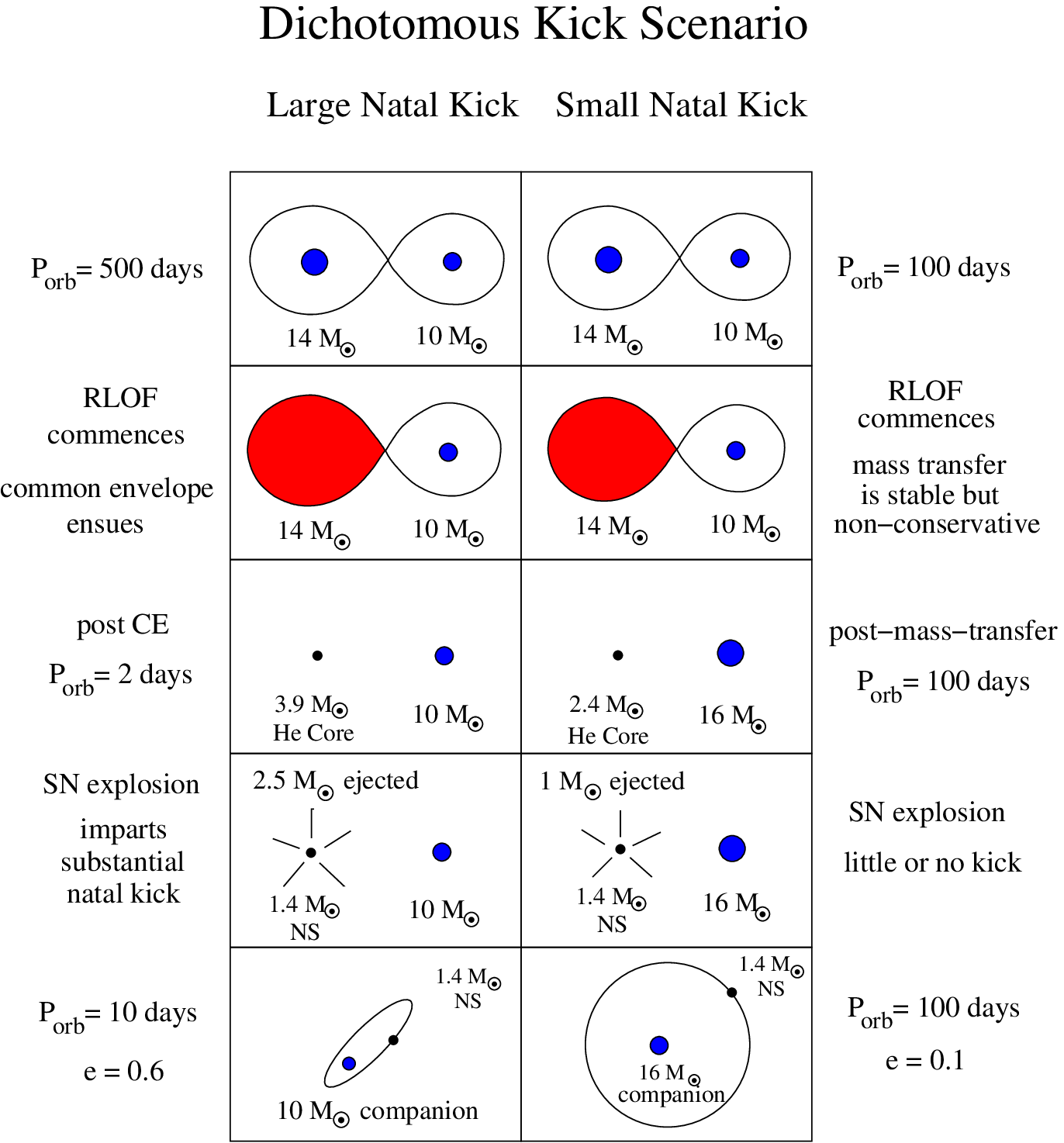,width=8cm}
\caption{Illustrative binary evolution scenarios leading to the
formation of high-mass X-ray binaries.  In the left panel, the binary
is sufficiently wide at the onset of mass transfer that the primary
has a fully formed $3.9\Ms$ He core.  By contrast, in the right panel
the primary's envelope is lost while the He core is substantially
lower in mass ($2.4\Ms$).  In the proposed dichotomous kick scenario
the former case leads to a large natal kick, while the latter case
results in a ``prompt (fast)'' supernova event and a smaller kick (see
text).}\label{fig:vdH}
\end{center}
\end{figure}
Given some critical mass ratio, which we take to be $q_c =0.5$, the
mass transfer from the primary to the secondary is assumed to be
stable if $q_i > q_c$ {\em and\/} the envelope of the primary is mostly
radiative when mass transfer begins (so-called Cases ``A'', ``early
B'', and ``early C''), and dynamically unstable if $q_i < q_c$ {\em
or} the primary has a convective envelope (Cases ``late B'' and ``late
C'').  In any case, we assume that the entire hydrogen-rich envelope
of the primary is removed during mass transfer, leaving only the
primary's hydrogen-depleted core.  

For stable mass transfer, we suppose that the secondary accretes a
fraction, $\beta = 0.75$, of material donated by the primary, and that
the remaining mass escapes the system with a specific angular momentum
that is $\alpha = 1.5$ times the orbital angular momentum per unit
reduced mass.  The value of $\alpha = 1.5$ is characteristic of mass
loss through the L2 point, but the choice of $\beta = 0.75$ is fairly
arbitrary -- although it is convenient, as it gives $a_f/a_i \sim 1$,
where $a_f$ is the final orbital separation.  Dynamically unstable
mass transfer results in a common-envelope phase, wherein the
secondary spirals into the envelope of the primary.  The
common-envelope phase was treated in the same way as in Pfahl et al.\
(2002c).  A fraction, $\eta_{\rm CE} \la 1$, of the initial orbital
energy is available to unbind the common envelope from the system.  If
{\em in}sufficient energy is available, the two stars will merge.
Otherwise, the envelope of the primary is dispersed, the secondary
emerges near the ZAMS without having accreted any significant amount
of mass, and the orbital separation is $\sim$100 times smaller.  A
merger results in nearly every case where $q_i < q_c$ and the
primary's envelope is radiative when mass transfer starts.  A number
of these evolutionary steps are illustrated in the top six panels of
the schematic shown in Figure~\ref{fig:vdH}.

Upon exhausting its remaining nuclear fuel, the exposed core of the
primary explodes as a Type Ib or Ic SN.  We assume that the supernova
mass loss and natal NS kick are instantaneous, and that the
orientations of the kicks are distributed isotropically.  In light of
the discussion above, the specific prescription adopted for NS kicks
is as follows.  Kick speeds are drawn from a Maxwellian distribution,
with a mean that depends on the initial mass of the primary and its
evolutionary state when it first fills its Roche lobe (see Fig.\
\ref{fig:vdH}).  If $8 < M_{1i} < 14 M_\odot$ and mass transfer begins
before helium ignition in the primary's core (Cases A and B), the mean
kick speed takes a small value of $30~{\rm km~s}^{-1}$, while in all
other cases the mean is $300~{\rm km~s}^{-1}$.  The range of masses
and the actual mean kick speeds used here are reasonable choices, but
are somewhat arbitrary, and were chosen mainly for illustrative purposes.

The results of our BPS calculation are shown in Figure~\ref{fig:BPS}
in the $P_{\rm orb} - M_2$ and $P_{\rm orb} - e$ planes, where $M_2$
is the mass of the secondary star, including the mass it has accreted
from the primary, and $e$ is the orbital eccentricity.  Each small dot
represents an incipient high-mass X-ray binary immediately after the
supernova explosion that produces the neutron star.  The red dots
represent systems which experienced the nominal natal neutron star
kicks (see also left column of Figure~\ref{fig:vdH}) while the blue
dots are systems where the primary lost its envelope early enough so
that the natal kicks were much smaller -- in accord with the
dichotomous kick scenario proposed herein (see right hand panel of
Fig.\ \ref{fig:vdH}).

The vertical blue strip in the $P_{\rm orb} - M_2$ plane of Fig.\
\ref{fig:BPS} results mostly from early Case B systems in which the
primaries had largely radiative envelopes and start mass transfer
before core He ignition.  The resultant mass transfer is stable and
the H-exhausted core evolves to a ``prompt (fast)'' SN explosion with
a small natal kick. These systems generally acquire small
eccentricities, especially for $P_{\rm orb}$ in the range of 10\,--\,100
days.  These are presumably the systems that evolve to become the
newly identified class of HMXBs with relatively wide, low-eccentricity
orbits (e.g., X Per/4U 0352 + 30; $\gamma$ Cas/MX 0053+604; XTE
J0543-568; the filled circles in Fig.~2).  For the systems of this
type with still wider orbits, even the small natal kicks assumed here
are sufficient to induce substantial eccentricities.

The blue ``tail''-shaped region in the $P_{\rm orb} - M_2$ plane of
Figure~\ref{fig:BPS} results from late Case B systems commencing mass
transfer when the primary had a large, convective envelope, but still
before core He core ignition.  These lead to common-envelope events
and result in systems with short orbital periods (i.e., $P_{\rm orb} <
10$ days; see e.g., Fig.\ \ref{fig:vdH}).  Since such short period
systems would, in any case, be circularized by tidal friction, it may
be difficult to distinguish these from their Case C cousins where the
primaries were much more evolved at the start of mass transfer, and
which received larger natal kicks.  Systems in these latter categories
(late Case B and late Case C) presumably include many of the famous
HMXBs such as Cen X-3, LMC X-4, 4U 0900-40, and SMC X-1, with
current-epoch (approximately) circular orbits and short orbital
periods.  The red dots in Figure~\ref{fig:BPS} are those receiving a
large kick, where the primary was highly evolved at the onset of mass
loss (Case C).  These likely account for the class of eccentric HMXB
systems with $P_{\rm orb}$ in the range of $\sim 20-200$ days, e.g.,
4U 0115+63, GX 301-2, 4U 1145-619, and EXO 2030+375
\citep{Bildsten1997} (filled triangles in Fig.~2) and the three radio
pulsars with massive companions in highly eccentric systems (filled
squares in Fig.~2) PSR 1259-63 (Johnston et al.\ 1992), PSR J0045-7319
(Kaspi et al.\ 1994, 1996) and PSR J1740-3052 (Stairs et al.\ 2001).

Finally, we note that this dichotomous kick scenario helps the
long-standing ``retention problem'' for neutron stars in globular
clusters.  In an earlier work \citep{Pfahl2002b} we reported that our
original dichotomous kick scenario yielded an enhancement factor of
about 4 over the fraction of retained neutron stars without the
appropriate subset of small kicks.  We also pointed out
\citep{Pfahl2002b} that a dichotomous kick scenario would likely
increase the theoretical formation rate of double neutron star systems
by approximately an order of magnitude.  Both of these enhancements
are expected to be carried over to this newer kick scenario.

\section{Discussion and future work}

As we have shown, the presence in a binary can dramatically affect the
structure of the core of a massive star at the time of core collapse.
Stars above $\sim 11\Ms$ are generally expected to have smaller iron
cores if they lose their envelopes in a close binaries.
Stars in the range of 8\,--\,11\Ms\ may explode in an
electron-capture supernova if they are in a close binary, while single
stars or stars in wide binaries will experience a second dredge-up
phase and are more likely to end their evolution as ONeMg white dwarfs. We
suggested that in the case of small iron cores or in the case of an
electron-capture supernova, the supernova occurs through a prompt (fast)
explosion rather than a delayed neutrino-driven explosion and argued
that this is more likely to produce neutron stars with low kick
velocities.

While speculative at the moment, this scenario has important
observational implications and suggests the need for futher
theoretical studies. These include a systematic exploration of the
late evolution of the cores of stars around 10\Ms, both for stars
evolved in isolation and in a close binary, the dependence on
metallicity and the amount of convective overshooting (e.g.,
\citealt{Poelarends2004,Eldridge2004}), and a re-examination of the
core-collapse phase for electron-capture supernovae and low-mass iron
cores.  One immediate prediction of this scenario is that neutron
stars that formed from ONeMg or low-mass iron cores should produce
neutron stars of systematically lower mass ($\sim 1.25\Ms$ for a
typical equation of state\footnote{The pulsar that formed last in the
recently discovered double-pulsar binary PSR J0737-3039 [Burgay et
al.\ 2003; Lyne et al.\ 2004] has a low mass of $\sim 1.25\Ms$,
although at present it is not clear whether the formation of the
second pulsar is consistent with a low-velocity kick.}).
\citet{Wanajo2003} suggested that prompt explosions are the site for
the r-process; if some of this r-processed matter is captured by the
companion stars, this may produce detectable chemical anomalies in
neutron-star companions (e.g., X Per) similar to the case of the
companion in Nova Scorpii
\citep{Israelian1999,Podsiadlowski2002}. Note also that, in this case,
only small amounts of iron and other heavy elements are ejected in the
supernova. If all neutron stars in globular clusters were to form
through such a channel, one would not expect significant chemical
pollution of the cluster with heavy elements from supernova ejecta
\citep{Price1993}.

Our scenario also has important implications for the minimum mass for
neutron-star formation; it suggests that only single stars more
massive than $\sim 10$\,--12\Ms\ become neutron stars, while in
binaries the mass may be as low as 6\,--\,8\Ms. The exact values will
depend both on the amount of convective overshooting and the
metallicity of the population; larger amounts of overshooting and
lower metallicity are both expected to shift these critical masses to
lower values. The detection of both young, single massive white dwarfs
and neutron stars in binaries in open clusters with a turnoff mass
around 10\Ms\ would provide direct evidence for such a dichotomy.
Another promising method to observationally constrain the minimum mass
for a core-collapse supernova is through the detection of the
progenitors of Type II-P supernovae, stars that have not lost all of
their H-rich envelopes at the time of the supernova explosion (see
e.g., \citealt{Smartt2002}).

Finally, it is worth mentioning another channel which is likely to
produce an electron-capture supernova and possibly a neutron star with
a low kick if the basic picture described in this paper is correct:
this involves the accretion-induced collapse (AIC) of a white dwarf to
form a neutron star, either because an ONeMg white dwarf accreting
from a companion star is pushed above the Chandraskekhar limiting mass
\citep{Nomoto1987b,Nomoto1991} or as the result of the merger of two
CO white dwarfs which is also likely to lead to the formation of a
neutron star rather than a Type Ia supernova \citep{Nomoto1985}.  In
principle, AIC could produce single low-velocity neutron stars with a
rate as high as $3\times 10^{-3}\,$yr$^{-1}$
\citep{Han1998,Nelemans2001}.

\acknowledgments We thank H.-Th.\ Janka and C.~F. Fryer for very useful
discussions. This work has been in part supported by the
Netherlands Organization for Scientific Research (NWO) (AJP), NASA ATP
grant NAG5-12522 (SR), a Chandra Fellowship held at the Chandra X-ray
center through grant PF2-30024 (EP) and the Department of Energy
under contract W-7405-ENG-36 (AH). AH has also been supported in part
by the Department of Energy under grant B341495 to the Center for
Astrophysical Thermonuclear Flashes at the University of Chicago.

\phantom{\citep{a,b,c,d,e,f,g}}

\clearpage





\begin{thebibliography}{}

\bibitem[Arzoumanian, Chernoff, \& Cordes(2002)]{Arzoumanian2002}
Arzoumanian, Z., Chernoff, D. F., \& Cordes, J. M. 2002, ApJ, 568, 289

\bibitem[Bildsten et al.(1997)]{Bildsten1997}
Bildsten, L.~et al.\ 1997, \apjs, 113, 367

\bibitem[Brandt \& Podsiadlowski(1995)]{Brandt1995}
Brandt, W. N., \& Podsiadlowski, Ph., 1995, MNRAS, 274, 461

\bibitem[Brown, Lee, \& Bethe(1999)]{Brown1999}
Brown, G. E., Lee, C.-H., Bethe, H. A. 1999, NewA, 4, 313

\bibitem[Brown et al.(2001)]{Brown2001}
Brown, G. E., Heger, A., Langer, N., Lee, C.-H., Wellstein, S., \&
Bethe, H. 2001, New Astronomy, 6, 457

\bibitem[(2999)]{} Burgay, M. 2003, Nature, 426, 531 

\bibitem[Cordes \& Chernoff(1998)]{Cordes1998}
Cordes, J. M., \& Chernoff, D. F. 1998, ApJ, 505, 315

\bibitem[Drukier(1996)]{Drukier1996} Drukier, G. A. 1996, MNRAS, 280, 498

\bibitem[Eldridge \& Tout(2004)]{Eldridge2004}
Eldridge, J., \& Tout, C. A. 2004, submitted

\bibitem[Fryer(1999)]{Fryer1999}
Fryer, C.~L. 1999, ApJ, 522, 413

\bibitem[Fryer(2003)]{Fryer2003}
Fryer, C.~L. 2003, in Stellar Collapse, Fryer, C.~L, ed.\ (Dordrecht,
Kluwer)


\bibitem[Fryer et al.(2002)]{Fryer2002}
Fryer, C. L., Heger, A., Langer, N., \& Wellstein, S. 2002,
ApJ, 578, 335

\bibitem[Fryer \& Kalogera(2001)]{Fryer2001}
Fryer, C.~L., \& Kalogera, V. 2001, ApJ, 554, 548

\bibitem[Fryer \& Warren(2002)]{Fryer2002b}
Fryer, C. L., \& Warren, M.~S. 2002, ApJ, 574, L65

\bibitem[a(2002)]{a} Fryer, C. L., \& Warren, M.~S. 2003, ApJ, in press (astro-ph/0309539)

\bibitem[Garc\'\i a-Berro, Ritossa, \& Iben(1997)]{Berro1997}
Gac\'\i a-Berro, E., Ritossa, C., Iben, I., Jr. 1997, ApJ, 485, 765

\bibitem[Habets(1986)]{Habets1986}
Habets, G. M. H. J. 1986, A\&A, 167, 61

\bibitem[Han(1998)]{Han1998}
Han, Z. 1998, MNRAS, 296, 1019

\bibitem[Han, Podsiadlowski, \& Eggleton(1994)]{Han1994}
Han, Z., Podsiadlowski, Ph., Eggleton, P. P. 1994, MNRAS, 270, 121

\bibitem[Hansen \& Phinney(1997)]{Hansen1997}
Hansen, B. M. S., \& Phinney, E. S. 1997, MNRAS, 291, 569

\bibitem[b(2002)]{b} Hartman J. M. 1997, A\&A, 322, 127

\bibitem[Heger \& Woosley(2003)]{Heger2003b}
Heger, A., \& Woosley, S. E. 2003, in Proc.\ Woods Hole 2001, Gamma-Ray
Bursts and Afterglow Astronomy (AIP), in press

\bibitem[Heger et al.(2003)]{Heger2003}
Heger, A., Woosley, S. E., Langer, N., \& Spruit, H. C. 2003,
in Maeder, A., \& Eenens, Ph., eds, 
Stellar Rotation, IAU Symp.\ No.\ 215, in press (astro-ph/0301374) 

\bibitem[Heger et al.(2002)]{Heger2002}
Heger, A., Woosley, S.~E., Rauscher, T., Hoffman, R.~D., \& Boyes, M.~M.
2002, New Astronomy Review, 46, 463


\bibitem[Hillebrandt, Nomoto, \& Wolff(1984)]{Hillebrandt1984}
Hillebrandt, W., Nomoto, K, \& Wolff, G. 1984, A\&A, 133, 175

\bibitem[Iben(1974)]{Iben1974}
Iben, I., Jr. 1974, ARA\&A, 12, 214

\bibitem[Iben, Ritossa, \& Garc\'\i a-Berro(1997)]{Iben1997}
Iben, I. Jr., Ritossa, C., \& Garc\'\i a-Berro, E. 1997, ApJ, 489, 772

\bibitem[Israelian et al.(1999)]{Israelian1999} Israelian, G., Rebolo,
R., Basri, G., Casares, J., \& Martin E. L. 1999, Nat, 401, 142

\bibitem[Ivanova \& Podsiadlowski(2003)]{Ivanova2003} Ivanova, N., \&
Podsiadlowski, Ph.\ 2003, in Hillebrandt, W., \& Leibundgut, B., eds,
From Twilight to Highlight: the Physics of Supernovae (Springer,
Berlin), p. 19



\bibitem[Janka et al.(2003)]{Janka2002} Janka, H.-T., Buras, R., Kifonidis,
K., \& Rampp, M. 2003,  in Hillebrandt, W., \& Leibundgut, B., eds,
From Twilight to Highlight: the Physics of Supernovae (Springer,
Berlin), p. 39

\bibitem[Johnston et al.(1992)]{johnston92} Johnston, S.,
Manchester, R.~N., Lyne, A.~G., Bailes, M., Kaspi, V.~M., Qiao, G., \&
D'Amico, N.\ 1992, \apjl, 387, L37

\bibitem[c(2002)]{c} Joss, P. C., \& Becker, J. A. 2003, in Hillebrandt, 
 W., \& Leibundgut, B., eds, From Twilight to Highlight: 
 the Physics of Supernovae (Springer, Berlin), p. 104

\bibitem[Kaspi et al.(1996)]{kaspi96} Kaspi, V.\ M., Bailes,
M., Manchester, R.\ N., Stappers, B.\ W., \& Bell, J.\ F.\ 1996, \nat, 
381, 584

\bibitem[Kaspi et al.(1994)]{kaspi94} Kaspi, V.~M., Johnston,
S., Bell, J.~F., Manchester, R.~N., Bailes, M., Bessell, M., Lyne, 
A.~G., \& D'Amico, N.\ 1994, \apjl, 423, L43

\bibitem[d(2002)]{d} Katz, J. I. 1975, Nat, 253, 698

\bibitem[e(2002)]{e} Katz, J. I. 1983, A\&A, 128, L1

\bibitem[Kippenhahn \& Weigert(1967)]{Kippenhahn1967}
Kippenhahn, R., \& Weigert, A. 1967, Z. Astrophys., 65, 251

\bibitem[Langer, Wellstein, \& Petrovic(2003a)]{Langer2003a} Langer,
N., Wellstein, \& Petrovic J., 2003, in {A massive star odyssey: from
main sequence to supernova}, IAU Symp.\ 212, ed.\
K. A. van der Hucht, A. Herrero, \& C. Esteban (San Francisco, ASP), p.~275

\bibitem[Langer et al.(2003)]{Langer2003b} Langer N., Yoon S.-C.,
Petrovic J., Heger A., 2003, in {Stellar Rotation}, IAU Symp.\ 215,
ed. A. Maeder \& P. Eenens (San Francisco: ASP), in press

\bibitem[Lauterborn(1970)]{Lauterborn1970}
Lauterborn, D. 1970, A\&A, 7, 150

\bibitem[(2999d)]{} Lyne, A. G., et al.\ 2004, Science, in press
(astro-ph/0401086)
 
\bibitem[Lyne \& Lorimer(1994)]{Lyne1994}
Lyne, A. G., \& Lorimer, D. R. 1994, Nature, 369, 127


\bibitem[Nelemans et al.(2001)]{Nelemans2001}
Nelemans, G., Yungelson, L. R., Portegies Zwart, S. F., \& Verbunt, F.
2001, A\&A 365 491

\bibitem[Nomoto(1984)]{Nomoto1984}
Nomoto, K. 1984, ApJ, 277, 791

\bibitem[Nomoto(1987a)]{Nomoto1987a}
Nomoto, K. 1987a, ApJ, 322, 206

\bibitem[Nomoto(1987b)]{Nomoto1987b}
Nomoto, K. 1987b, in IAU Symp.\ 125, The Origin and Evolution of Neutron
Stars, ed. D. J. Helfand \& J.-H. Huang (Dordrecht: Kluwer), p. 281

\bibitem[Nomoto \& Iben(1985)]{Nomoto1985}
Nomoto, K., \& Iben, I., Jr. 1985, 297, 531

\bibitem[Nomoto \& Kondo(1991)]{Nomoto1991}
Nomoto, K., \& Kondo, Y. 1991, ApJ 367, L19

\bibitem[Paczy\'nski(1976)]{Paczynski1976}
Paczy\'nski, B. 1976, in Structure and Evolution of Close Binary Systems,
ed.\ P. P. Eggleton, S. Mitton, \& J. Whelan (Dordrecht: Reidel), p. 75

\bibitem[Pfahl, Rappaport, \& Podsiadlowski(2002a)]{Pfahl2002a}
Pfahl, E., Rappaport, S., \& Podsiadlowski Ph. 2002a, ApJ, 571, L37 

\bibitem[Pfahl, Rappaport, \& Podsiadlowski(2002b)]{Pfahl2002b}
Pfahl, E., Rappaport, S., \& Podsiadlowski Ph.\ 2002b, ApJ, 573, 283 

\bibitem[Pfahl et al.(2002c)]{Pfahl2002c} 
Pfahl, E., Rappaport, S., Podsiadlowski, Ph., \& Spruit, H. 2002c,
ApJ, 574, 364  

\bibitem[Pfahl, Rappaport, \& Podsiadlowski(2003)]{Pfahl2003} 
Pfahl, E., Rappaport, S., \& Podsiadlowski, Ph. 2003,
ApJ, 597, 1036 

\bibitem[Podsiadlowski, Joss, \& Hsu(1992)]{Podsiadlowski1992}
Podsiadlowski, Ph., Joss, P. C., \& Hsu, J. J. L. 1992, ApJ, 391, 246

\bibitem[Podsiadlowski, et al.(2002)]{Podsiadlowski2002}
Podsiadlowski, Ph., Nomoto, K., Maeda, K., Nakamura, T., Mazzali, P.,
\& Schmidt, B. 2002, ApJ, 567, 491

\bibitem[Podsiadlowski, Rappaport, \& Han(2003)]{Podsiadlowski2003}
Podsiadlowski, Ph., Rappaport, S., \& Han, Z. 2003, MNRAS, 341, 385


\bibitem[Poelarends \& Langer(2004)]{Poelarends2004}
Poelarends, A. J. T., \& Langer N. 2004, in preparation

\bibitem[Pols \& Dewi(2002)]{Pols2002}
Pols, O. R, \& Dewi, J. D. W. 2002, Publ.\ Astron.\ Soc.\ Australia, 19, 233

\bibitem[Pols et al.(1997)]{Pols1997} Pols O. R., Tout C. A.,
Schr\"oder K.-P., Eggleton P. P., Manners, J., 1997, MNRAS, 289, 869

\bibitem[Price \& Podsiadlowski(1993)]{Price1993} 
Price, N. M., \& Podsiadlowski, Ph., 1993, in {ASP Conference Series,
Vol. 48, The Globular Cluster -- Galaxy Connection}, ed.\ G. H. Smith
\& J. P. Brodie (ASP, San Francisco), p. 721

\bibitem[Ritossa, Garc\'\i a-Berro, \& Iben(1996)]{Ritossa1996}
Ritossa, C., Garc\'\i a-Berro, E., Iben, I., Jr. 1996, ApJ, 460, 489

\bibitem[Ritossa, Garc\'\i a-Berro, \& Iben(1999)]{Ritossa1999}
Ritossa, C., Garc\'\i a-Berro, E., Iben, I., Jr. 1999, ApJ, 515, 381

 \bibitem[f(2002)]{f} Scheck, L., Plewa, T., Janka, H.-Th., Kifonidis, K., 
\& M\"uller, E. 2004, PhRvL, 92, 1103

\bibitem[Schr\"oder et al.(1997)]{Schroeder1997} Schr\"oder K.-P.,
Pols O. R., Eggleton P. P., 1997, MNRAS, 285, 696

\bibitem[Smartt et al.(2002)]{Smartt2002}
Smartt, S. J., Gilmore, G. F., Tout, C. A., \& Hodgin, S. T. 2002,
ApJ, 565, 1089

\bibitem[Spruit \& Phinney(1998)]{Spruit1998} Spruit, H. C., Phinney,
E. S. 1998, Nature, 393, 139
\bibitem[Sumiyoshi et al.(2001)]{Sumiyoshi2001}
Sumiyoshi, K., Terasawa, M., Matthews, G. J., Kajino, T., Yamada, S., 
\& Suzuki, H. 2001, ApJ, 562, 880

\bibitem[Stairs et al.(2001)]{stairs01} Stairs, I.~H.~et al.\
2001, \mnras, 325, 979

\bibitem[Wanajo et al.(2003)]{Wanajo2003}
Wanajo, S., Tamamura, M., Itoh, N., Nomoto, K., Ishimaru, Y., Beers, T. C.,
Nozawa, S. 2003, ApJ, 593, 968

\bibitem[Wellstein \& Langer(1999)]{Wellstein1999}
Wellstein, S., \& Langer, N. 1999, A\&A, 350, 148

\bibitem[Wellstein, Langer, \& Braun(2001)]{Wellstein2001}
Wellstein, S., Langer, N., Braun, H. 2001, A\&A, 369, 939


\bibitem[Woosley, Langer, \& Weaver(1995a)]{Woosley1995a}
Woosley, S.~E., Langer, N., \& Weaver, T. A. 1995, ApJ, 448, 315

\bibitem[Woosley \& Weaver(1995)]{Woosley1995b}
Woosley, S. E., \& Weaver, T. A. 1995, ApJS, 101, 181

\end{thebibliography}
\end{document}